\documentclass[12pt]{article}
\usepackage{amsmath}
\newlength{\extraspace}
\newlength{\extraspaces}
\newcommand{\be}{\begin{equation}
\addtolength{\abovedisplayskip}{\extraspaces}
\addtolength{\belowdisplayskip}{\extraspaces}
\addtolength{\abovedisplayshortskip}{\extraspace}
\addtolength{\belowdisplayshortskip}{\extraspace}}
\newcommand{\ee}{\end{equation}}
\newcommand{\ba}{\begin{eqnarray}
\addtolength{\abovedisplayskip}{\extraspaces}
\addtolength{\belowdisplayskip}{\extraspaces}
\addtolength{\abovedisplayshortskip}{\extraspace}
\addtolength{\belowdisplayshortskip}{\extraspace}}
\newcommand{\ea}{\end{eqnarray}}
\newcommand{\nonu}{\nonumber \\[.5mm]}
\newcommand{\A}{&\!\!\!}
\begin{document}
\thispagestyle{empty}
\begin{flushright}
\end{flushright}
\vspace{7mm}
%
%
\begin{center}
{\large{\bf Phase Transition of NLSUSY Space-Time \\[1mm]
 and \\[2mm]
Unity of Nature }} 
\footnote{
\tt Talk given by K. Shima at {\it Scale and Duality in Quantum Science Workshop},
04-06 November, 2009, RIMS-YITP, Kyoto University, Kyoto.
} \\ [20mm]
%
{\sc Kazunari Shima}
\footnote{
\tt e-mail: shima@sit.ac.jp} \ 
and \ 
{\sc Motomu Tsuda}
\footnote{
\tt e-mail: tsuda@sit.ac.jp} 
\\[5mm]
{\it Laboratory of Physics, 
Saitama Institute of Technology \\
Fukaya, Saitama 369-0293, Japan} \\[20mm]
\begin{abstract}
The mysterious relation between the large scale structure of the universe 
and the tiny (Planck) scale structure of the particle physics, 
e.g. the observed mysterious relation 
between the (dark) energy density (and the dark matter) of the universe 
and the origin of the tiny neutrino mass (and the SUSY breaking mass scale) of the particle physics 
may be explained simply by the nonlinear supersymmmetric general relativity theory (NLSUSY GR). 
\\[5mm]
\noindent
%
\noindent
\end{abstract}
\end{center}

\newpage
\noindent
\section{Introduction}
Supersymmetry (SUSY) and its spontaneous breakdown are profound notions related to {the space-time symmetry}, 
therefore, to be studied not only in the low energy particle physics 
but also in the cosmology, i.e. in the framework necessarily accomodating graviton as well. 
We found that $SO(10)$ super-Poincar\'e (sP) group accomodates minimally the standard model (SM) 
with just three generations 
of quarks and leptons and the graviton as the low energy physical states in the 
{\it single} irreducible representation of $SO(10)$ sP \cite{KS0}. 
We have decomposed $10$ supercharges as 
${\underline {10}}_{SO(10)}={\underline 5}_{SU(5)}+{\underline 5}^{*}_{SU(5)}$ 
according to $SO(10) \supset SU(5)$ 
and assigned the same quantum numbers as those of $\underline 5$ of $SU(5)$ GUT to ${\underline 5}_{SU(5)}$ 
satisfying $Q_{e}=I_{z}+{1 \over 2}(B-L)$. 
Regarding ${\underline 5}_{SU(5)}$ as a quintet of hypothetical spin-${1 \over 2}$ constituents ({\it superon}) 
for all observed particles we have proposed the {\it superon-quintet model} (SQM) of matter, 
which may give potentially simple explanations about  the proton stability, various mixings of states, etc., 
though qualitative so far \cite{KS}. 
We discuss  in this article the field theoretical description of SQM including gravity, 
called {\it superon-graviton model} (SGM), 
and show some particle physics consequences of SGM. 
The physical and mathematical (geometrical) origin of the (mass) scale and 
the relation between the Planck scale (gravity dominating) physics and the low energy particle physics 
(the duality of SGM theory) are discussed in some detail. 
The familiar SUSY SM can be regarded as a equivalent low energy theory of SQM in asymptotic flat space-time,  
which may indicate the relativistic second order phase transition of massless SGM dictated 
by the global structure (symmetry) of space-time (graviton).

\section{Nonlinear supersymmetric general relativity \\
(NLSUSY GR)}
For the  the field theoretical description  of SQM including gravity, 
the supersymmetric coupling of spin-${1 \over 2}$ objects (superon) 
with spin-$2$ graviton is necessary.  
Nonlinear supersymmetric general relativity theory (NLSUSY GR) \cite{KSa}, 
which is based upon the general relativity (GR) principle 
and the nonlinear (NL) representation \cite{VA} of supersymmetry (SUSY) \cite{WZ,GL}, 
is the simple model and proposes a new paradigm called the SGM scenario \cite{KS,KSa,ST1,ST2} 
for the unified description of space-time and matter beyond (behind) the (SUSY) SM. 
In NLSUSY GR,  new (generalized) space-time, {\it SGM space-time} \cite{KS}, is introduced, 
where tangent space-time has the NLSUSY structure, 
i.e. flat tangent space-time is specified not only by the $SO(3,1)$ Minkowski coodinates $x_a$ 
but also by $SL(2,C)$ Grassmann coordinates $\psi^i_\alpha$ ($i = 1, 2, \cdots, N$) for NLSUSY. 
The Grassmann coordinates in  new (SGM) space-time are 
coset space coordinates of ${super GL(4,R) \over GL(4,R)}$ 
which allows to interpret $\psi$ as the NG-fermions superon 
associated with the spontaneous breaking of super-$GL(4,R)$ down to $GL(4,R)$. 
The fundamental action of NLSUSY GR is obtained in the Einstein-Hilbert (EH) form in SGM space-time 
by extending the geometrical arguments of GR in Riemann space-time, 
which has a priori the spontaneous SUSY breaking and $SO(N)$ sP symmetry \cite{ST1}. 
In order to see the abovementioned particle physics consequences of  NLSUSY GR, i.e. 
{\it the relation between the large scale structure of space-time and the low energy particle physics}, 
we start with the fundamental NLSUSY GR action of EH-type in new (SGM) space-time given by \cite{KSa} 
\begin{equation}
L_{\rm NLSUSYGR}(w) =-{c^4 \over {16 \pi G}} \vert w \vert \{ \Omega(w) + \Lambda \}, 
\label{NLSUSYGR}
\end{equation}
where $G$ is the Newton gravitational constant, $\Lambda>0$ is a ({\it small}) cosmological constant, 
$\Omega(w)$ is the unified Ricci scalar curvature 
in terms of the unified vierbein $w^a{}_\mu(x)$ of new space-time(and the inverse $w^\mu{}_a$) defined by 
\begin{equation}
w^a{}_\mu = e^a{}_\mu + t^a{}_\mu(\psi), 
\ \ 
t^a{}_\mu(\psi) = {\kappa^2 \over 2i} 
(\bar\psi^i \gamma^a \partial_\mu \psi^i - \partial_\mu \bar\psi^i \gamma^a \psi^i), 
\label{unified-w}
\end{equation}
and $\vert w \vert = \det w^a{}_\mu$. 
In Eq.(\ref{unified-w}), $e^a{}_\mu$ is the ordinary vierbein of GR for the local $SO(3,1)$, 
$t^a{}_\mu(\psi)$ is the stress-energy-momentum tensor (i.e. the mimic vierbein) 
of the NG fermion $\psi^i(x)$ for the local $SL(2,C)$ 
and $\kappa$ is an arbitrary constant of NLSUSY with the dimemsion (mass)$^{-2}$. 
Note that $e^a{}_\mu$ and $t^a{}_\mu(\psi)$ contribute equally to the curvature of space-time, 
which may be regarded as the Mach's principle in ultimate space-time. 

The NLSUSY GR action (\ref{NLSUSYGR}) possesses promissing large symmetries 
isomorphic to $SO(N)$ sP group \cite{ST1}; 
namely, $L_{\rm NLSUSYGR}(w)$ is invariant under \\[2mm]
%
[{\rm new \ NLSUSY}] $\otimes$ [{\rm local \ $GL(4,R)$}] 
$\otimes$ [{\rm local \ Lorentz}] $\otimes$ [{\rm local \ spinor \ translation}] \\
\hspace*{2.5cm} 
$\otimes$ [{\rm global}\ $SO(N)$] $\otimes$ [{\rm local}\ $U(1)^N$] $\otimes$ [{\rm Chiral}]. \\[2mm]
%
%
%
%
Note that the no-go theorem is overcome (circumvented) in a sense that 
the nontivial $N$-extended SUSY gravity theory with $N > 8$ has been constructed by using NLSUSY, 
i.e. by the vacuum (flat space-time) degeneracy. 
%
%
%
\section{Linearization of NLSUSY and \\ 
Low Energy Particle Physics of NLSUSY GR}
New {\it empty} (SGM) space-time for {\it everything} described by the ({\it vacuum}) 
EH-type NLSUSY GR action (\ref{NLSUSYGR}) is unstable due to NLSUSY structure of tangent space-time 
and decays spontaneously (called Big Decay) to ordinary Riemann space-time with the NG fermions (superon matter) 
called superon-graviton model (SGM). 
The SGM action is given by the following; 
\begin{equation}
L_{\rm SGM}(e,\psi) = -{c^4 \over {16 \pi G}} e \vert w_{\rm VA} \vert \{ R(e) + \Lambda + T(e, \psi)\} 
\label{SGM}
\end{equation}
where $R(e)$ is the ordinary Ricci scalar curvature of ordinary EH action, 
$T(e,\psi)$ represents highly nonlinear gravitational interaction terms of $\psi^i$, 
and $\vert w_{\rm VA} \vert = \det w^a{}_b = \det (\delta^a_b + t^a{}_b)$ 
is the invariant volume of NLSUSY \cite{VA}. 
Remarkablly the cosmological term in $L_{\rm NLSUSYGR}(w)$ of Eq.(\ref{NLSUSYGR}) 
(i.e. the constant energy density of ultimate space-time) 
reduces to the NLSUSY action \cite{VA}, $L_{\rm NLSUSY}(\psi) = -{1 \over {2 \kappa^2}} \vert w_{\rm VA} \vert$, 
in Riemann-flat $e_a{}^\mu(x) \rightarrow \delta_a^\mu$ space-time, 
i.e. the arbitrary constant $\kappa$ of NLSUSY is now fixed to 
\begin{equation}
\kappa^{-2} = {{c^4 \Lambda} \over {8 \pi G}} 
\label{kappa}
\end{equation}
in SGM scenario. 
Note that the NLSUSY GR action (\ref{NLSUSYGR}) and the SGM action (\ref{SGM}) possess 
different asymptotic flat space-time, i.e. 
SGM-flat $w_a{}^\mu \rightarrow \delta_a^\mu$ space-time 
and Riemann-flat $e_a{}^\mu \rightarrow \delta_a{}^\mu$ space-time, respectively. 
The spontaneous symmetry breaking (SSB) in NLSUSY GR (called Big Decay of new space-time) 
produces a fundamental mass scale 
depending on the $\Lambda$ and $G$ through the relation (\ref{kappa}), 
whose effect survives as the evidence of SGM scenario in the (low energy) particle physics 
in asymptotic flat space-time, as shown below. 
Remember that the potential of $L_{\rm NLSUSYGR}(w)$ and $L_{\rm SGM}(e,\psi)$ (massless superon-graviton) 
is $\sim {\Lambda \over G}>0$.%

To see the (low energy) particle physics content in asymptotic Riemann-flat
($e^a{}_\mu \rightarrow \delta^a_\mu$) space-time 
we discuss  $N=2$ SUSY in two dimensional space-time for simplicity, 
for $N=2$ in the SGM scenario gives the minimal and realistic $N = 2$ LSUSY model \cite{STT}. 
By  the systematic arguments for $N=2$ SUSY theory  the equivalencee  
between  $N=2$  NLSUSY model and $N=2$ LSUSY QED theory is demonstrated \cite{ST3,ST4}; namely, 
\begin{equation}
L_{N=2{\rm SGM}}(e,\psi) \overset{e^a{}_\mu \rightarrow \delta^a_\mu}{\longrightarrow} 
L_{N=2{\rm NLSUSY}}(\psi) 
= L_{N=2{\rm SUSYQED}}({\bf V},{\bf \Phi}) + [{\rm tot.\ der.\ terms}], 
\label{NLSUSY-SUSYQED}
\end{equation}
which we call the NL/L SUSY relation (in flat spae-time).
In the relation (\ref{NLSUSY-SUSYQED}), the $N = 2$ NLSUSY action $L_{N=2{\rm NLSUSY}}(\psi)$ 
for the two (Majorana) NG-fermions superon $\psi^i$ $(i = 1, 2)$ is written in $d = 2$ as follows; 
\begin{eqnarray}
&\!\!\! &\!\!\! 
L_{N=2{\rm NLSUSY}}(\psi) 
\nonumber \\
&\!\!\! &\!\!\! 
\hspace*{5mm} = -{1 \over {2 \kappa^2}} \vert w_{\rm VA} \vert 
= - {1 \over {2 \kappa^2}} 
\left\{ 1 + t^a{}_a + {1 \over 2!}(t^a{}_a t^b{}_b - t^a{}_b t^b{}_a) 
\right\} 
\nonumber \\
&\!\!\! &\!\!\! 
\hspace*{5mm} = - {1 \over {2 \kappa^2}} 
\bigg\{ 1 - i \kappa^2 \bar\psi^i \!\!\not\!\partial \psi^i 
- {1 \over 2} \kappa^4 
( \bar\psi^i \!\!\not\!\partial \psi^i \bar\psi^j \!\!\not\!\partial \psi^j 
- \bar\psi^i \gamma^a \partial_b \psi^i \bar\psi^j \gamma^b \partial_a \psi^j ) 
\bigg\}, 
\label{NLSUSYaction}
\end{eqnarray}
where $\kappa$ is a constant with the dimension $({\rm mass})^{-1}$, 
which satisfies the relation (\ref{kappa}). 

While, the $N = 2$ LSUSY QED action $L_{N=2{\rm SUSYQED}}({\bf V},{\bf \Phi})$ in Eq.(\ref{NLSUSY-SUSYQED})
consists of a $N = 2$ {\it minimal} off-shell vector supermultiplet ${\bf V}$, 
a $N = 2$ off-shell scalar supermultiplet denoted ${\bf \Phi}$, a Fayet-Iliopoulos (FI) $D$ term 
and Yukawa interactions.  
The explicit component form of $L_{N=2{\rm SUSYQED}}({\bf V},{\bf \Phi})$ in $d = 2$ 
with $U(1)_{local}$
for the massless case is given ; 
\begin{eqnarray}
L_{N=2{\rm SUSYQED}}({\bf V},{\bf \Phi}) 
&\!\!\! = &\!\!\!- {1 \over 4} (F_{ab})^2 
+ {i \over 2} \bar\lambda^i \!\!\not\!\partial \lambda^i 
+ {1 \over 2} (\partial_a A)^2 
+ {1 \over 2} (\partial_a \phi)^2 
+ {1 \over 2} D^2 
- {\xi \over \kappa} D 
\nonumber \\[.5mm]
& & 
+ {i \over 2} \bar\chi \!\!\not\!\partial \chi 
+ {1 \over 2} (\partial_a B^i)^2 
+ {i \over 2} \bar\nu \!\!\not\!\partial \nu 
+ {1 \over 2} (F^i)^2 
\nonumber \\[.5mm]
& & 
+ f ( A \bar\lambda^i \lambda^i + \epsilon^{ij} \phi \bar\lambda^i \gamma_5 \lambda^j 
- A^2 D + \phi^2 D + \epsilon^{ab} A \phi F_{ab} ) 
\nonumber \\[.5mm]
& & 
+ e \bigg\{ i v_a \bar\chi \gamma^a \nu 
- \epsilon^{ij} v^a B^i \partial_a B^j 
+ \bar\lambda^i \chi B^i 
+ \epsilon^{ij} \bar\lambda^i \nu B^j 
\nonumber \\[.5mm]
& & 
- {1 \over 2} D (B^i)^2 
+ {1 \over 2} A (\bar\chi \chi + \bar\nu \nu) 
- \phi \bar\chi \gamma_5 \nu \bigg\}
\nonumber \\[.5mm]
& & 
+ {1 \over 2} e^2 (v_a{}^2 - A^2 - \phi^2) (B^i)^2, 
\label{SQEDaction}
\end{eqnarray}
where $(v^a, \lambda^i, A, \phi, D)$ ($F_{ab} = \partial_a v_b - \partial_b v_a$) 
and  ($\chi$, $B^i$, $\nu$, $F^i$)  are the component fields of the minimal off-shell supermultiplets 
${\bf V}$ and  ${\bf \Phi}$, respectively. 
Also $\xi$ in the FI $D$ term is an arbitrary dimensionless parameter turning to a magnitude of SUSY breaking mass, 
and $f$ and $e$ are Yukawa and gauge coupling constants with the dimension (mass)$^1$ (in $d = 2$). 
The $N = 2$ LSUSY QED action (\ref{SQEDaction}) can be rewritten in the familiar manifestly covariant form 
by using the superfield formulation (for further details see Ref.\cite{ST4}). 

From the mathematical scientific nature of this workshop, 
we would like to discuss the NL/L SUSY relation in detail. \\
The equivalence (NL/L SUSY relation) of the two theories (\ref{NLSUSY-SUSYQED}) means; 
(i) the component fields of $({\bf V},{\bf \Phi})$ in the $N = 2$ LSUSY QED action (\ref{SQEDaction}) are 
expressed explicitly as composites of the NG fermions $\psi^i$ (called {\it SUSY invariant relations}) 
which terminate at ${\cal O}\{ (\psi^i)^{4} \}$ (for the $d = 2$, $N = 2$ case), 
(ii) the familiar LSUSY transformations on  $({\bf V},{\bf \Phi})$ are precisely reproduced 
by the NLSUSY transformations on the constituents and (iii) substituting the SUSY invariant relations 
into $L_{\rm LSUSYQED}$, we confirm  (\ref{NLSUSY-SUSYQED}). 

Consider the superfields on specific supertranslations 
of superspace coordinates \cite{IK,UZ} with a parameter $\zeta^i = - \kappa \psi^i$, 
which are denoted by $(x'^a, \theta_\alpha'^i)$, 
\ba
\A \A 
x'^a = x^a + i \kappa \bar\theta^i \gamma^a \psi^i, 
\nonu
\A \A 
\theta'^i = \theta^i - \kappa \psi^i. 
\ea
%
We define the $N = 2$ general gauge and the $N = 2$ scalar (matter) superfields 
on $(x'^a, \theta_\alpha'^i)$ as 
\be
{\cal V}(x', \theta') \equiv \tilde{\cal V}(x, \theta, \psi), 
\ \ \Phi^i(x', \theta') \equiv \tilde \Phi^i(x, \theta, \psi), 
\label{SFpsi}
\ee
where $\tilde{\cal V}(x, \theta, \psi)$ and $\tilde \Phi^i(x, \theta, \psi)$ 
may be expanded as 
\ba
\tilde{\cal V}(x, \theta, \psi) 
\A = \A \tilde C(x) + \bar\theta^i \tilde\Lambda^i(x) 
+ {1 \over 2} \bar\theta^i \theta^j \tilde M^{ij}(x) 
- {1 \over 2} \bar\theta^i \theta^i \tilde M^{jj}(x) 
+ {1 \over 4} \epsilon^{ij} \bar\theta^i \gamma_5 \theta^j \tilde\phi(x) 
\nonu
\A \A 
- {i \over 4} \epsilon^{ij} \bar\theta^i \gamma_a \theta^j \tilde v^a(x) 
- {1 \over 2} \bar\theta^i \theta^i \bar\theta^j \tilde\lambda^j(x) 
- {1 \over 8} \bar\theta^i \theta^i \bar\theta^j \theta^j \tilde D(x), 
\label{tVSF} \\
\tilde \Phi^i(x, \theta, \psi) 
\A = \A \tilde B^i(x) + \bar\theta^i \tilde \chi(x) - \epsilon^{ij} \bar\theta^j \tilde \nu(x) 
- {1 \over 2} \bar\theta^j \theta^j \tilde F^i(x) + \bar\theta^i \theta^j \tilde F^j(x) + \cdots. 
\label{tSSF}
\ea
In Eqs.(\ref{tVSF}) and (\ref{tSSF}) the component fields 
$\tilde\varphi_{\cal V}^I(x) = \{ \tilde C(x), \tilde\Lambda^i(x), \cdots \}$ 
and $\tilde\varphi_\Phi^I(x) = \{ \tilde B^i(x), \tilde\chi(x), \cdots \}$ 
are functionals of the initial component fields $\varphi_{\cal V}^I(x)$ and $\varphi_\Phi^I(x)$ 
in $({\bf V},{\bf \Phi})$  and the NG fermions. 
Take the supertranslation on $x', \theta'$ we have 
\be
\delta_\zeta \tilde{\cal V}(x, \theta, \psi) = \xi^a \partial_a \tilde{\cal V}(x, \theta. \psi), 
\ \ \delta_\zeta \tilde \Phi^i(x, \theta, \psi) = \xi^a \partial_a \tilde \Phi^i(x, \theta, \psi) 
\ee
with $\xi^a = i \kappa \bar\psi^i \gamma^a \zeta^i$, 
which mean that the components $\tilde\varphi_{\cal V}^I(x)$ 
and $\tilde\varphi_\Phi^I(x)$ do not transform each other 
and that the constant values of the tilded fields are conserved.  
Therefore, the following SUSY invariant constraints can be imposed, 
\ba
\label{SUSYconst-VSF}\tilde\varphi_{\cal V}^I(x) = {\rm constant}, 
\ \ \tilde\varphi_\Phi^I(x) = {\rm constant}, 
\ea
%
%
which are invariant (conserved quantities) under the supertrasformations. 
The most general form of the SUSY invariant constraints are as follows; 
\ba
\A \A 
\tilde C = \xi_c, \ \ \tilde\Lambda^i = \xi_\Lambda^i, 
\ \ \tilde M^{ij} = \xi_M^{ij}, \ \ \tilde\phi = \xi_\phi, 
\ \ \tilde v^a = \xi_v^a, \ \ \tilde\lambda^i = \xi_\lambda^i, 
\ \ \tilde D = {\xi \over \kappa}, 
\label{SUSYconst-VSF1}
\\
\A \A 
\tilde B^i = \xi_B^i, \ \ \tilde\chi = \xi_\chi, \ \ \tilde\nu = \xi_\nu, 
\ \ \tilde F^i = {\xi^i \over \kappa}, 
\label{SUSYconst-SSF1}
\ea
Solving the SUSY invariant constraints for 
the initial component fields $\varphi_{\cal V}^I$ and $\varphi_\Phi^I$, 
we obtain the following 
SUSY invariant relations $\varphi_{\cal V}^I = \varphi_{\cal V}^I(\psi)$ and $\varphi_\Phi^I = \varphi_\Phi^I(\psi)$, 
%
\ba
C \A = \A \xi_c + \kappa \bar\psi^i \xi_\Lambda^i 
+ {1 \over 2} \kappa^2 (\xi_M^{ij} \bar\psi^i \psi^j - \xi_M^{ii} \bar\psi^j \psi^j) 
+ {1 \over 4} \xi_\phi \kappa^2 \epsilon^{ij} \bar\psi^i \gamma_5 \psi^j 
- {i \over 4} \xi_v^a \kappa^2 \epsilon^{ij} \bar\psi^i \gamma_a \psi^j 
\nonu
\A \A 
- {1 \over 2} \kappa^3 \bar\psi^i \psi^i \bar\psi^j \xi_\lambda^j 
- {1 \over 8} \xi \kappa^3 \bar\psi^i \psi^i \bar\psi^j \psi^j, 
\nonu
\Lambda^i \A = \A \xi_\Lambda^i 
+ \kappa (\xi_M^{ij} \psi^j - \xi_M^{jj} \psi^i) 
+ {1 \over 2} \xi_\phi \kappa \epsilon^{ij} \gamma_5 \psi^j 
- {i \over 2} \xi_v^a \kappa \epsilon^{ij} \gamma_a \psi^j 
\nonu
\A \A 
- {1 \over 2} \xi_\lambda^i \kappa^2 \bar\psi^j \psi^j 
+ {1 \over 2} \kappa^2 
(\psi^j \bar\psi^i \xi_\lambda^j 
- \gamma_5 \psi^j \bar\psi^i \gamma_5 \xi_\lambda^j 
- \gamma_a \psi^j \bar\psi^i \gamma^a \xi_\lambda^j) 
\nonu
\A \A 
- {1 \over 2} \xi \kappa^2 \psi^i \bar\psi^j \psi^j 
- i \kappa \!\!\not\!\partial C(\psi) \psi^i, 
\nonu
M^{ij} \A = \A \xi_M^{ij} 
+ \kappa \bar\psi^{(i} \xi_\lambda^{j)} 
+{1 \over 2} \xi \kappa \bar\psi^i \psi^j 
+ i \kappa \epsilon^{(i \vert k \vert} \epsilon^{j)l} \bar\psi^k \!\!\not\!\partial \Lambda^l(\psi) 
- {1 \over 2} \kappa^2 \epsilon^{ik} \epsilon^{jl} \bar\psi^k \psi^l \partial_a \partial^a C(\psi), 
\nonu
\phi \A = \A \xi_\phi 
- \kappa \epsilon^{ij} \bar\psi^i \gamma_5 \xi_\lambda^j 
- {1 \over 2} \xi \kappa \epsilon^{ij} \bar\psi^i \gamma_5 \psi^j 
- i \kappa \epsilon^{ij} \bar\psi^i \gamma_5 \!\!\not\!\partial \Lambda^j(\psi) 
+ {1 \over 2} \kappa^2 \epsilon^{ij} \bar\psi^i \gamma_5 \psi^j \partial_a \partial^a C(\psi), 
\nonu
v^a \A = \A \xi_v^a 
- i \kappa \epsilon^{ij} \bar\psi^i \gamma^a \xi_\lambda^j 
- {i \over 2} \xi \kappa \epsilon^{ij} \bar\psi^i \gamma^a \psi^j 
- \kappa \epsilon^{ij} \bar\psi^i \!\!\not\!\partial \gamma^a \Lambda^j(\psi) 
+ {i \over 2} \kappa^2 \epsilon^{ij} \bar\psi^i \gamma^a \psi^j \partial_b \partial^b C(\psi) 
\nonu
\A \A 
- i \kappa^2 \epsilon^{ij} \bar\psi^i \gamma^b \psi^j \partial^a \partial_b C(\psi), 
\nonu
\lambda^i \A = \A \xi_\Lambda^i 
+ \xi \psi^i - i \kappa \!\!\not\!\partial M^{ij}(\psi) \psi^j 
+ {i \over 2} \kappa \epsilon^{ab} \epsilon^{ij} \gamma_a \psi^j \partial_b \phi(\psi) 
\nonu
\A \A 
- {1 \over 2} \kappa \epsilon^{ij} \left\{ \psi^j \partial_a v^a(\psi) 
- {1 \over 2} \epsilon^{ab} \gamma_5 \psi^j F_{ab}(\psi) \right\} 
\nonu
\A \A
- {1 \over 2} \kappa^2 \{ \partial_a \partial^a \Lambda^i(\psi) \bar\psi^j \psi^j 
- \partial_a \partial^a \Lambda^j(\psi) \bar\psi^i \psi^j 
- \gamma_5 \partial_a \partial^a \Lambda^j(\psi) \bar\psi^i \gamma_5 \psi^j 
\nonu
\A \A 
- \gamma_a \partial_b \partial^b \Lambda^j(\psi) \bar\psi^i \gamma^a \psi^j 
+ 2 \!\!\not\!\partial \partial_a \Lambda^j(\psi) \bar\psi^i \gamma^a \psi^j \} 
- {i \over 2} \kappa^3 \!\!\not\!\partial \partial_a \partial^a C(\psi) \psi^i \bar\psi^j \psi^j, 
\nonu
D \A = \A {\xi \over \kappa} - i \kappa \bar\psi^i \!\!\not\!\partial \lambda^i(\psi) 
\nonu
\A \A 
+ {1 \over 2} \kappa^2 \left\{ \bar\psi^i \psi^j \partial_a \partial^a M^{ij}(\psi) 
- {1 \over 2} \epsilon^{ij} \bar\psi^i \gamma_5 \psi^j \partial_a \partial^a \phi(\psi) \right. 
\nonu
\A \A 
\left. 
+ {i \over 2} \epsilon^{ij} \bar\psi^i \gamma_a \psi^j \partial_b \partial^b v^a(\psi) 
- i \epsilon^{ij} \bar\psi^i \gamma_a \psi^j \partial_a \partial_b v^b(\psi) \right\} 
\nonu
\A \A
- {i \over 2} \kappa^3 \bar\psi^i \psi^i \bar\psi^j \!\!\not\!\partial \partial_a \partial^a \Lambda^j(\psi) 
+ {1 \over 8} \kappa^4 \bar\psi^i \psi^i \bar\psi^j \psi^j (\partial_a \partial^a)^2 C(\psi), 
\label{SUSYrelation-VSF}
\ea
and 
%
\ba
B^i \A = \A \xi_B^i + \kappa (\bar\psi^i \xi_\chi - \epsilon^{ij} \bar\psi^j \xi_\nu) 
- {1 \over 2} \kappa^2 \{ \bar\psi^j \psi^j F^i(\psi) - 2 \bar\psi^i \psi^j F^j(\psi) 
+ 2 i \bar\psi^i \!\!\not\!\partial B^j(\psi) \psi^j \} 
\nonu
\A \A 
- i \kappa^3 \bar\psi^j \psi^j \{ \bar\psi^i \!\!\not\!\partial \chi(\psi) 
- \epsilon^{ik} \bar\psi^k \!\!\not\!\partial \nu(\psi) \} 
+ {3 \over 8} \kappa^4 \bar\psi^j \psi^j \bar\psi^k \psi^k \partial_a \partial^a B^i(\psi), 
\nonu
\chi \A = \A \xi_\chi + \kappa \{ \psi^i F^i(\psi) - i \!\!\not\!\partial B^i(\psi) \psi^i \} 
\nonu
\A \A 
- {i \over 2} \kappa^2 [ \not\!\partial \chi(\psi) \bar\psi^i \psi^i 
- \epsilon^{ij} \{ \psi^i \bar\psi^j \!\!\not\!\partial \nu(\psi) 
- \gamma^a \psi^i \bar\psi^j \partial_a \nu(\psi) \} ] 
\nonu
\A \A 
+ {1 \over 2} \kappa^3 \psi^i \bar\psi^j \psi^j \partial_a \partial^a B^i(\psi) 
+ {i \over 2} \kappa^3 \!\!\not\!\partial F^i(\psi) \psi^i \bar\psi^j \psi^j 
+ {1 \over 8} \kappa^4 \partial_a \partial^a \chi(\psi) \bar\psi^i \psi^i \bar\psi^j \psi^j, 
\nonu
\nu \A = \A \xi_\nu - \kappa \epsilon^{ij} \{ \psi^i F^j(\psi) - i \!\!\not\!\partial B^i(\psi) \psi^j \} 
\nonu
\A \A 
- {i \over 2} \kappa^2 [ \not\!\partial \nu(\psi) \bar\psi^i \psi^i 
+ \epsilon^{ij} \{ \psi^i \bar\psi^j \!\!\not\!\partial \chi(\psi) 
- \gamma^a \psi^i \bar\psi^j \partial_a \chi(\psi) \} ] 
\nonu
\A \A 
+ {1 \over 2} \kappa^3 \epsilon^{ij} \psi^i \bar\psi^k \psi^k \partial_a \partial^a B^j(\psi) 
+ {i \over 2} \kappa^3 \epsilon^{ij} \!\!\not\!\partial F^i(\psi) \psi^j \bar\psi^k \psi^k 
+ {1 \over 8} \kappa^4 \partial_a \partial^a \nu(\psi) \bar\psi^i \psi^i \bar\psi^j \psi^j, 
\nonu
F^i \A = \A {\xi^i \over \kappa} - i \kappa \{ \bar\psi^i \!\!\not\!\partial \chi(\psi) 
+ \epsilon^{ij} \bar\psi^j \!\!\not\!\partial \nu(\psi) \} 
\nonu
\A \A 
- {1 \over 2} \kappa^2 \bar\psi^j \psi^j \partial_a \partial^a B^i(\psi) + \kappa^2 \bar\psi^i \psi^j \partial_a \partial^a B^j(\psi) 
+ i \kappa^2 \bar\psi^i \!\!\not\!\partial F^j(\psi) \psi^j 
\nonu
\A \A 
+ {1 \over 2} \kappa^3 \bar\psi^j \psi^j \{ \bar\psi^i \partial_a \partial^a \chi(\psi) 
+ \epsilon^{ik} \bar\psi^k \partial_a \partial^a \nu(\psi) \} 
- {1 \over 8} \kappa^4 \bar\psi^j \psi^j \bar\psi^k \psi^k \partial_a \partial^a F^i(\psi). 
\label{SUSYrelation-SSF}
\ea
For simplicity we adopt the following simple SUSY invariant constraints 
(i.e. reductions of the auxiliary fields only) 
\ba
\A \A 
\tilde C = \xi_c, \ \tilde\Lambda^i = \tilde M^{ij} = \tilde\phi = \tilde v^a = \tilde\lambda^i = 0, 
\ \tilde D = {\xi \over \kappa}, \ \tilde B^i = \tilde\chi = \tilde\nu = 0, \ \tilde F^i = {\xi^i \over \kappa}, 
\label{SUSYconst-VSF2}
\ea
accordingly the SUSY invariant relations (\ref{SUSYrelation-VSF}) and (\ref{SUSYrelation-SSF}) reduce to 
\ba
C \A = \A \xi_c - {1 \over 8} \xi \kappa^3 \bar\psi^i \psi^i \bar\psi^j \psi^j \vert w \vert, 
\nonu
\Lambda^i \A = \A - {1 \over 2} \xi \kappa^2 
\psi^i \bar\psi^j \psi^j \vert w \vert, 
\nonu
M^{ij} \A = \A {1 \over 2} \xi \kappa \bar\psi^i \psi^j \vert w \vert, 
\nonu
\phi \A = \A - {1 \over 2} \xi \kappa \epsilon^{ij} \bar\psi^i \gamma_5 \psi^j \vert w \vert, 
\nonu
v^a \A = \A - {i \over 2} \xi \kappa \epsilon^{ij} \bar\psi^i \gamma^a \psi^j \vert w \vert, 
\nonu
\lambda^i \A = \A \xi \psi^i \vert w \vert, 
\nonu
D \A = \A {\xi \over \kappa} \vert w \vert, 
\label{SUSYrelation-VSF1}
\end{eqnarray}
and 
\ba
\chi \A = \A \xi^i \left[ \psi^i \vert w \vert
+ {i \over 2} \kappa^2 \partial_a 
( \gamma^a \psi^i \bar\psi^j \psi^j \vert w \vert 
) \right], 
\nonu
B^i \A = \A - \kappa \left( {1 \over 2} \xi^i \bar\psi^j \psi^j 
- \xi^j \bar\psi^i \psi^j \right) \vert w \vert, 
\nonu
\nu \A = \A \xi^i \epsilon^{ij} \left[ \psi^j \vert w \vert 
+ {i \over 2} \kappa^2 \partial_a 
( \gamma^a \psi^j \bar\psi^k \psi^k \vert w \vert 
) \right], 
\nonu
F^i \A = \A {1 \over \kappa} \xi^i \left\{ \vert w \vert 
+ {1 \over 8} \kappa^3 
\partial_a \partial^a ( \bar\psi^j \psi^j \bar\psi^k \psi^k \vert w \vert ) 
\right\} 
\nonu
\A \A 
- i \kappa \xi^j \partial_a ( \bar\psi^i \gamma^a \psi^j \vert w \vert ), 
\label{SUSYrelation-SSF1}
\end{eqnarray}
which are written in the form containing some vanishing terms due to $(\psi^i)^5 \equiv 0$. 
Further by adopting the Wess-Zumino gauge,  auxiliary fields $C, \Lambda^i, M^{12}, M^{11} - M^{22}$ are 
gauged away. 
By substituting the SUSY invariant relations for the remaining minimal supermultiplet 
(with $A \equiv M^{ii}$) 
into $L_{N=2 {\rm LSUSYQED}}$ we obtain $L_{N=2 {\rm NLSUSY}}$, i.e. NL/L SUSY relation (\ref{NLSUSY-SUSYQED}) 
for $N=2$ SUSY is established.

Now we briefly show the (physical) vacuum structure of $N = 2$ LSUSY QED action (\ref{SQEDaction}) 
related (equivalent) to the $N = 2$ NLSUSY action (\ref{NLSUSYaction}) \cite{STL,STLa}. 
The vacuum is determined by the minimum of the potential $V(A, \phi, B^i, D)$ in the action (\ref{SQEDaction}).    
%
%
The potential is  given by using the equation of motion for the auxiliary field $D$ as 
\begin{equation}
V(A, \phi, B^i) = {1 \over 2} f^2 \left\{ A^2 - \phi^2 + {e \over 2f} (B^i)^2 
+ {\xi \over {f \kappa}} \right\}^2 + {1 \over 2} e^2 (A^2 + \phi^2) (B^i)^2 \ge 0, 
\label{potential}
\end{equation}
The configurations of the fields corresponding to vacua $\delta V(A, \phi, B^i)=0$ 
in $(A, \phi, B^i)$-space in the potential (\ref{potential}), 
which are dominated by $SO(1,3)$ or $SO(3,1)$ isometries, 
are classified according to the signs of the parameters $e, f, \xi, \kappa$. 

By adopting the simple parametrization $(\rho, \theta, \varphi, \omega)$ for the vacuum configuration 
of $(A, \phi, B^i)$-space  and expanding the fields $(A, \phi, B^i)$ around the vacua 
we obtain the particle (mass) spectra of the linearized theory, $N = 2$ LSUSY QED. 
We have found the  vacuum $V(A, \phi, B^i)=0$ in the $SO(3,1)$ isometry \cite{STL,STLa} 
which describes $N=2$ LSUSY QED possessing the similar structure to the lepton sector of the SM.\\[2mm]
%
%
%
\hspace*{30mm} 
one charged Dirac fermion ($\psi_D{}^c \sim \chi + i \nu$), \\
\hspace*{30mm} 
one neutral (Dirac) fermion ($\lambda_D{}^0 \sim \lambda^1 - i \lambda^2$), \\
\hspace*{30mm} 
one massless vector (a photon) ($v_a$), \\
\hspace*{30mm} 
one charged scalar ($\phi^c \sim \theta + i \varphi$), \\
\hspace*{30mm} 
one neutral complex scalar ($\phi^0 \sim \rho {}(+ i \omega)$), \\[2mm]
with masses $m_{\phi^0}^2 = m_{\lambda_D{}^0 }^2 = 4 f^2 k^2 = -{{4 \xi f} \over \kappa}$,  
$m_{\psi_D{}^c} ^2 = m_{\phi^c}^2 = e^2 k^2 =-{{\xi e^2} \over {\kappa f}}$, 
$m_{v_{a}} = m_{\omega} = 0$,
%
which are the composites of NG-fermions superon 
and the vacuum 
breaks SUSY alone  spontaneously (The local $U(1)$ is not broken. $\omega$ is a NG boson 
of SSB of $SU(2)_{global}$ and disappears provided the corresponding 
local gauge symmetry is introduced as in the SM.) 
Remarkably these arguments show that the true vacuum $V=0$ of (asymptotic flat space-time of) 
$L_{N=2{\rm SGM}}(e,\psi)$ 
is achieved by the compositeness of fields (eigenstates) of the supermultiplet
of {\it global} $N = 2$ LSUSY QED. 
This phenomena may be regarded as the {\it relativistic} second order phase transition 
of massless superon-graviton NLSUSY system, 
which is dictated by the symmetry of space-time (analogous to the superconducting states achieved 
by the Cooper pair 
of electron-phonon system).

As for the cosmological significances of $N = 2$ SUSY QED in the SGM scenario, 
the (physical) vacuum for the above model explains (predicts) simply the observed mysterious (numerical) relation 
between {\it the (dark) energy density of the universe} $\rho_D$ ($\sim {{c^4 \Lambda} \over {8 \pi G}}$) 
and {\it the neutrino mass} $m_\nu$, 
\begin{equation}
\rho_D^{\rm obs} \sim (10^{-12} GeV)^4 \sim (m_\nu){}^4 
\sim {\Lambda \over G} \ (\sim {g_{\rm sv}}^2), 
\label{myst-relation}
\end{equation}
provided $- \xi f \sim O(1)$ and $\lambda_D{}^{0}$ is identified with  the neutrino, 
which gives a new insight into the origin of (small) mass \cite{ST2,STL,STLa} and 
produce the mass hielarchy by the factor ${e \over f}$($\sim O({m_{e} \over m_{\nu}})$ in case of $\psi_D{}^c$ as electron!). ($g_{\rm sv}$ is the superon-vacuum coupling constant via the supercurrent.)

Furthermore, the neutral scalar field ${\phi^{0} (\sim  \rho) }$ with mass $\sim  O( m_\nu)$ 
of the radial mode in the vacuum configuration may be a candidate of {\it the dark matter}, 
for $N = 2$ LSUSY QED structure and the radial mode in the vacuum are preserved in the realistic large $N$ SUSY GUT model.  
%
%
The no-go theorem for $N > 8$ SUSY may be overcome in a sense that the linearized (equivalent) $N>8$ LSUSY theory would be 
{\it massive} theory with SSB.
By taking the more {\it general} auxiliary-field structure $\xi_{c} \neq 0$ the NL/L SUSY relation gives
{\it the magnitude of the bare (dimensionless) gauge coupling constant $e$} 
(i.e. the fine structure constant $\alpha = {e^2 \over 4\pi}$)  
in terms of {\it vacuum values of auxiliary-fields} \cite{ST5}: 
\begin{equation}
e = {\ln({\xi^i{}^2 \over {\xi^2 - 1}}) \over 4 \xi_C}, 
%
\label{f-xi}
\end{equation}
where $e$ is the bare gauge coupling constant, $\xi$, $\xi^i$ and $\xi_C$ are the vacuum-values (parameters) 
of auxiliary-fields.  
This mechanism is natural and very favourable for SGM scenario as a theory for everything. 
%

\section{Conclusions}
We have proposed a new paradigm for describing the unity of nature, 
where the ultimate shape of nature is new unstable space-time 
described by the NLSUSY GR action $L_{\rm NLSUSYGR}(w)$ in the form of 
the free EH action for empty space-time with the constant energy density $\sim {\Lambda \over G}$. 
Big Decay of new space-time described by $L_{\rm NLSUSYGR}(w)$ creates 
ordinary Riemann space-time with {\it massless} spin-${1 \over 2}$ superon (matter) 
described by the SGM action $L_{\rm SGM}(e,\psi)$,  
which ignites the Big Bang of the universe accompanying the dark energy (cosmological constant). 
Interestingly on Riemann-flat tangent space (in the local frame), 
the familiar renormalizable LSUSY theory emerges on the true vacuum of SGM action $L_{\rm SGM}(e,\psi)$ 
as composite-eigenstates of superon. 
We have seen that the physics before/of the Big Bang is not the metaphysics but may play crucial roles 
for understanding unsolved problems of the universe and the particle physics, which can be tested. 
%
%
In fact, we have shown explicitly that the realistic $N=2$ LSUSY QED {\it composite} theory 
emerges in the true vacuum of $N=2$ NLSUSY theory on Minkowski tangent space-time and 
$N=3$ LSUSY YM {\it composite} theory \cite{ST6} from $N=3$ NLSUSY theory, as well, 
which gives new insights into the origin of mass and the cosmological problems.   
The cosmological implications of the composite SGM scenario seem promissing but deserve further studies. 
We can anticipate that NL/L SUSY relation holds for any $N$ SUSY and 
the physical cosequences obtained in $d=2$ hold in $d=4$ as well. 
%
The extension to large $N$, especially to $N = 5$ 
is important for {\it superon\ quintet\ hypothesis} of SGM scenario 
with ${N = \underline{10} = \underline{5}+\underline{5^{*}}}$ for equipping the $SU(5)$ GUT structure \cite{KS}  
and to  $N = 4$ may shed new light on the mathematical structures of 
the anomaly free non-trivial $d=4$ field theory. 
($N=10$ SGM predicts double-charge heavy lepton state $E^{2+}$ \cite{KS0}). 
%
%
%
Linearizing SGM action ${L_{\rm SGM}(e,\psi)}$ on curved space-time, 
which elucidates the topological structure of space-time \cite{SK}, is a challenge. 
The corresponding NL/L SUSY relation will give the supergravity (SUGRA) 
analogue with the vacuum breaking SUSY spontaneously. 
Locally homomorphic non-compact groups $SO(1,3)$ and $SL(2,C)$ for  
space-time degrees of freedom are analogues of compact groups $SO(3)$ and $SU(2)$ for gauge degrees of freedom  
of 't Hooft-Polyakov monopole.
The physical and mathematical meanings of the black hole as a singularity of space-time and 
the role of the equivalence principle are to be studied in detail in NLSUSY GR and SGM scenario. 
NLSUSY GR with extra space-time dimensions equipped with the Big Decay is also an interesting problem, 
which can give the framework for describing all observed particles as elementary {\it \`{a} la} Kaluza-Klein. 
%

Finally we speculate that $L_{SGM}$ describes the superfluidity of space-time  
%
and $\kappa^{-2}$ is the chemical potential of SGM space-time.

%
One of the authors (K.S) would like to thank Professor I. Ojima for the warm hospitality 
during the workshop at YITP and RIMS at Kyoto University. 
%

\newpage

%
\newcommand{\NP}[1]{{\it Nucl.\ Phys.\ }{\bf #1}}
\newcommand{\PL}[1]{{\it Phys.\ Lett.\ }{\bf #1}}
\newcommand{\CMP}[1]{{\it Commun.\ Math.\ Phys.\ }{\bf #1}}
\newcommand{\MPL}[1]{{\it Mod.\ Phys.\ Lett.\ }{\bf #1}}
\newcommand{\IJMP}[1]{{\it Int.\ J. Mod.\ Phys.\ }{\bf #1}}
\newcommand{\PR}[1]{{\it Phys.\ Rev.\ }{\bf #1}}
\newcommand{\PRL}[1]{{\it Phys.\ Rev.\ Lett.\ }{\bf #1}}
\newcommand{\PTP}[1]{{\it Prog.\ Theor.\ Phys.\ }{\bf #1}}
\newcommand{\PTPS}[1]{{\it Prog.\ Theor.\ Phys.\ Suppl.\ }{\bf #1}}
\newcommand{\AP}[1]{{\it Ann.\ Phys.\ }{\bf #1}}

\begin{thebibliography}{100}
%
\bibitem{KS0}
K. Shima, {\it Z. Phys. C} {\bf 18} (1983) 341.

\bibitem{KS}
K. Shima, {\it European Phys. J. C} {\bf 7} (1999) 341. 

\bibitem{KSa}
K. Shima, {\it Phys. Lett. B} {\bf 501} (2001) 237. 

\bibitem{VA}
D.V. Volkov and V.P. Akulov,  
{\it JETP Lett.} {\bf 16} (1972) 438; 
{\it Phys. Lett. B} {\bf 46} (1973) 109. 

\bibitem{WZ}
J. Wess and B. Zumino, {\it Phys. Lett. B} {\bf 49} (1974) 52. 

\bibitem{GL}
Yu. A. Golfand and E.S. Likhtman, {\it JETP Lett.} {\bf 13} (1971) 323. 

\bibitem{ST1}
K. Shima and M. Tsuda, {\it Phys. Lett. B} {\bf 507} (2001) 260; 
{\it Class. Quant. Grav.} {\bf 19} (2002) 5101. 

\bibitem{ST2}
K. Shima and M. Tsuda, {\it PoS HEP2005} (2006) 011. 

\bibitem{IK}
E.A. Ivanov and A.A. Kapustnikov, {\it J. Phys. A} {\bf 11} (1978) 2375; {\it J. Phys. G} {\bf 8} (1982) 167. 

\bibitem{Ro}
M. Ro\v{c}ek, {\it Phys. Rev. Lett.} {\bf 41} (1978) 451. 

\bibitem{UZ}
T. Uematsu and C.K. Zachos, {\it Nucl. Phys. B} {\bf 201} (1982) 250. 

\bibitem{STT}
K. Shima, Y. Tanii and M. Tsuda, {\it Phys. Lett. B} {\bf 525} (2002) 183; {\it ibid} {\bf 546} (2002) 162. 

%
\bibitem{ST3}
K. Shima and M. Tsuda, {\it Mod. Phys. Lett. A} {\bf 22} (2007) 1085; {\it ibid} {\bf 22} (2007) 3027. 

\bibitem{ST4}
K. Shima and M. Tsuda, {\it Phys. Lett. B} {\bf 666} (2008) 410; 
{\it Mod. Phys. Lett. A} {\bf 24} (2009) 185. 

\bibitem{STL}
K. Shima and M. Tsuda, {\it Phys. Lett. B} {\bf 645} (2007) 455. 

\bibitem{STLa} K. Shima, M. Tsuda and W. Lang, {\it Phys. Lett. B} {\bf 659} (2008) 741, 
Erratum {\it ibid} {\bf 660} (2008) 612, {\it ibid} {\bf 672} (2009) 413. 

\bibitem{ST5}
K. Shima and M. Tsuda, {\em Nuovo Cimento} {\bf 124B} (2009) 645. 

\bibitem{ST6}
K. Shima and M. Tsuda, {\it Phys. Lett. B} {\bf 687} (2010) 89. 

\bibitem{SK}
K. Shima and M. Kasuya,  {\em Phys. Rev. D} {\bf 22} (1980) 290. 

%
%
%
\end{thebibliography}
\end{document}